\documentclass[prl, superscriptaddress, notitlepage]{revtex4-1}
\usepackage{lipsum}
\usepackage{amsfonts}
\usepackage{amssymb}
\usepackage[english]{babel}
\usepackage{hyperref}   
\usepackage{color}
\usepackage{amsmath}
\usepackage{hyphenat}
\usepackage{graphicx}  
\usepackage{caption}

\newcommand{\pa}{\partial}

\newcommand{\bn}{\mbox{\boldmath $n$}}
\newcommand{\bmm}{\mbox{\boldmath $m$}}
\newcommand{\bx}{\mbox{\boldmath $x$}}

\newcommand{\bR}{\mbox{\boldmath $R$}}

\newcommand{\bpsi}{\mbox{\boldmath $\psi$}}
\newcommand{\bB}{\mbox{\boldmath $B$}}

\newcommand{\be}{\mbox{\boldmath $e$}}
\newcommand{\bp}{\mbox{\boldmath $p$}}
\newcommand{\bvarepsilon}{\mbox{\boldmath $\varepsilon$}}
\newcommand{\beginsupplement}{%
        \setcounter{table}{0}
        \renewcommand{\thetable}{S\arabic{table}}%
        \setcounter{figure}{0}
        \renewcommand{\thefigure}{S\arabic{figure}}%
     }
\begin{document}
\title{Composite Skyrmion bags in two-dimensional materials}

\author{David Foster}
\affiliation{H.~H.~Wills Physics Laboratory, University of Bristol, Bristol BS8 1TL, UK}
\author{Charles Kind}
\affiliation{School of Mathematics, University of Bristol, Bristol BS8 1TW, UK}
\author{Paul J. Ackerman}
\affiliation{Department of Physics and Soft Materials Research Center, University of Colorado, Boulder, Colorado 80309, USA}
\author{Jung-Shen B. Tai}
\affiliation{Department of Physics and Soft Materials Research Center, University of Colorado, Boulder, Colorado 80309, USA}
\author{Mark R.~Dennis}
\affiliation{H.~H.~Wills Physics Laboratory, University of Bristol, Bristol BS8 1TL, UK}
\affiliation{School of Physics and Astronomy, University of Birmingham, Birmingham B12 2TT, UK}
\author{Ivan I.~Smalyukh}
\affiliation{Department of Physics and Soft Materials Research Center, University of Colorado, Boulder, Colorado 80309, USA}
\affiliation{Department of Electrical, Computer, and Energy Engineering, Materials Science and Engineering Program, University of Colorado, Boulder, Colorado 80309, USA}
\affiliation{Renewable and Sustainable Energy Institute, National Renewable Energy Laboratory and University of Colorado, Boulder, Colorado 80309, USA}
\date{\today}

\begin{abstract}
\textbf{
Skyrmions are particle-like topological excitations, studied in various condensed matter systems and models of high-energy physics (HEP). 
They occur as stable spin textures in certain planar magnetic materials \cite{bogdanov1989thermodynamically,Cortes2017,muhlbauer2009skyrmion,yu2010real} and as configurations in chiral nematic liquid crystals \cite{Fuk}, having been originally proposed as model of atomic nuclei \cite{skyrme}.
Since magnetic Skyrmions can be accelerated with a current \cite{Woo2017,Iwasaki2013b,fert2013skyrmions}, they have the potential to encode bits in low-power magnetic storage devices \cite{Woo2016}.
Drawing on techniques from HEP, we demonstrate that magnetic and liquid crystal Skyrmions interact like orientation dependent, localised particles, explaining previously observed Skyrmion behaviour. 
This interaction motivates the construction of \emph{Skyrmion bags}: textures of high \emph{topological degree} which we realise experimentally in liquid crystals, and in magnetic materials by computer simulations.
These Skyrmion bags configurations are nested multiple Skyrmions, which act like single Skyrmions in pairwise interaction, and under the influence of a current in magnetic materials.
These results emphasize equivalent Skyrmion behaviour in different physical systems, and suggest new, high-density magnetic memory storage devices based on Skyrmion bags.
}
\end{abstract}
\maketitle
Topological invariants, such as the number of holes or windings in a texture, can be associated with the quantum numbers in a physical system.
This idea led T.H.R.~Skyrme, almost 60 years ago, to propose a topological model for atomic nuclei which has so-called \textit{Skyrmions} as minimum energy configurations which are localised in space and are stabilised by an integer valued topology number called the \textit{degree}, which corresponds to baryon number. 
Subsequently, Witten \cite{witten} derived the Skyrme model as a low energy effective model of quantum chromodynamics, in the large colour limit.
Skyrmions have also been observed in 2D magnetic materials, where $\bn(\bx)$ is the orientation of the magnetisation at the point $\bx$ \cite{bogdanov1989thermodynamically,Cortes2017,muhlbauer2009skyrmion,yu2010real}. 
Trains of magnetic Skyrmions have been proposed as a method of encoding data in racetrack memory technology \cite{Schulz2012,Yu2012,fert2013skyrmions,zhang2015magnetic} (potentially applicable to biological neural engineering \cite{brain}).
Skyrmions have also been experimentally observed in chiral nematic liquid crystal films \cite{Fuk}, which have a similar theoretical description to magnetic Skyrmions, where the liquid crystal molecules can be either polar vectors \cite{zhang2015ferromagnetic} (as the magnets) or directors, which do not distinguish $\pm \bn$. 
A full theory of Skyrmion-Skyrmion interaction in magnetic materials and liquid crystals, necessary to develop applications, has so far been lacking.

Here we build on the relationship between Skyrmions in the three different areas of physics.  
Extending a 2D calculation from High Energy Physics (HEP) \cite{Piette:1994ug,Foster:2015cpa}, we show that the interaction between magnetic and liquid crystal Skyrmions is a Yukawa type repulsion, consistent with a previously investigated numerical model \cite{lin2013particle}. 
We verify this in chiral nematic liquid crystals by experiment, and numerically with simulations of magnetic materials.
This interaction turns out to depend on the Skyrmions' relative orientation, explaining several previously observed phenomena. 
This leads us to investigate the viability of Skyrmion configurations with higher topological degree, which we call \emph{Skyrmion bags}. 
We develop a common mathematical description for the behaviour of Skyrmion bags, and in magnetic simulation and liquid crystal experiments show that these are stable and interact like single Skyrmions of large radius. 
From dynamical micro-magnetic simulations of bags in a racetrack device, we find magnetic Skyrmion bags can be driven by a current similar to single magnetic Skyrmions. 
This suggests the possibility of Skyrmion bags racetrack memory devices, which would be more robust and have a higher memory density than conventional Skyrmion racetrack memory.

\begin{figure}[!htb]
   \includegraphics[width=0.95\textwidth]{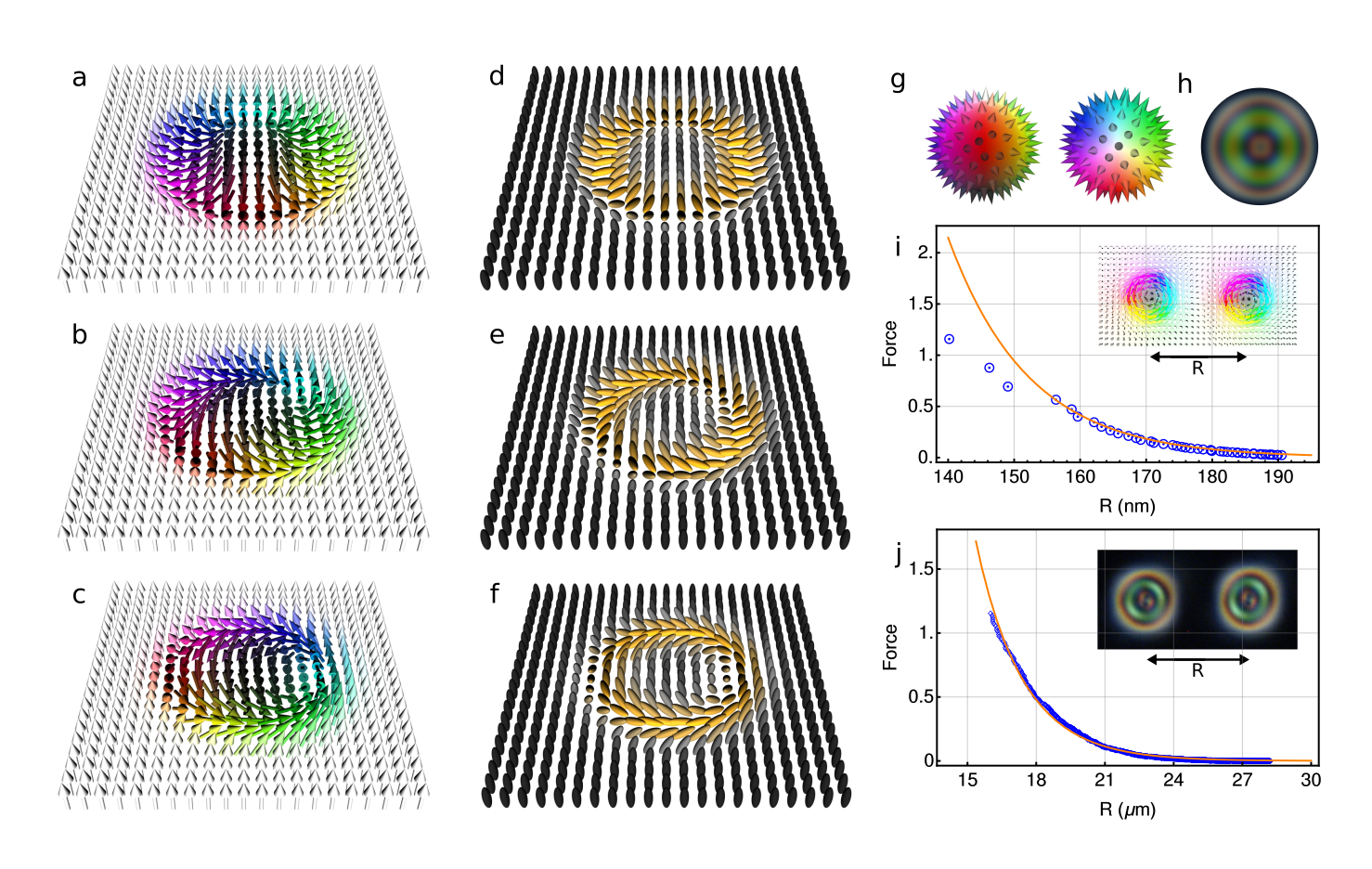}
   \caption{\textbf{Magnetic and liquid crystal Skyrmions.}
   \textbf{(a-c)} Three conformations of a degree one hedgehog Skyrmion \eqref{anz} with $\gamma=0,\pi/4,\pi/2$ respectively.
   Note that changing the internal orientation rotates the colours in the plane. In each case the set of arrows points in each direction on the 2-sphere once.
   \textbf{(d-f)} Three conformations of a hedgehog Skyrmion \eqref{anz} with $\gamma=0,\pi/4,\pi/2$ respectively for the chiral nematic liquid crystal director field. Here the directors do not have arrow heads, so opposite points on the 2-sphere are identified.
   \textbf{(g)} The colour scheme for the vectors in \textbf{(a-c)}, with hues and brightness determined by the position of $\bn$ on the $2$-sphere, viewing from the side and from above. 
   \textbf{(h)} Theoretical computer-simulated optical micrograph showing a liquid crystal Skyrmion in a polarizing optical microscope.
   \textbf{(i)} Dynamical MuMax3 simulation of two magnetic Skyrmions repelling (points), with the dipole interaction potential fitted \eqref{int_pot}. Typical Skyrmion diameter in this simulation regime is $110$ nm.
   \textbf{(j)} Experimental measurement of two liquid crystal Skyrmions repelling (points), with the dipole interaction potential fitted \eqref{int_pot}.
}
\label{Fig1}
\end{figure}

For an appropriate parameter regime, both chiral liquid crystals and magnetic configurations can be described by the energy functional
\begin{align}\label{eng}
E=\int \left(\frac{J}{2} \pa_i \bn \cdot \pa_i\bn+D\bn\cdot(\nabla \times \bn)\right)d^2 \bx + E_{\mathrm{ext}},
\end{align}
where $J$ and $D$ are coupling constants, and the second term is the Dzyaloshinskii\hyp{}Moriya interaction (DMI)\cite{nagaosa2013topological,dzyaloshinsky1958thermodynamic,moriya1960anisotropic,fert1980role}
(further discussion of the correspondence between magnetic and liquid crystals may be found in the Supplementary Information). 
For magnetic systems, $E_{\mathrm{ext}}$ is the integral over the plane of $-\bB\cdot \bn$, where $\bB=\mu \be$ is an external magnetic field, normal to the material. 
In chiral nematic liquid crystals, $E_{\mathrm{ext}}$ depends on the boundary conditions of the cell, which causes confinement and has a similar effect to a magnetic field.
Skyrmions are minimum energy configurations of the energy functional \eqref{eng}.
An important tool in understanding Skyrmions is the so-called \emph{hedgehog ansatz} \cite{nagaosa2013topological}, 
\begin{align} \label{anz}
\bn(\bx)=\left(\sin f(r)\cos(m\theta+\gamma),\sin f(r)\sin(m\theta+\gamma),\cos f(r) \right),
\end{align}
where $r^2=x^2+y^2$, $f(r)$ is some monotonic function satisfying $f(0)=\pi$, $f(\infty)=0$, $\theta$ is the polar angle of $\bn$ and $m\in \mathbb{Z}$ is the topological degree. 
$\gamma$ is the internal orientation of the Skyrmion (sometimes called the chirality). 
It is well understood \cite{melcher2014chiral,derrick1964comments} that the Skyrmion with $m=1, \gamma =-\frac{\pi}{2}$ is the minimiser over all homotopy classes.
Along with experimental observations \cite{kiselev2011chiral, muhlbauer2009skyrmion, yu2010real}, this shows that Skyrmions appear in collections of single-degree excitations. 
Thus in both magnetic systems and liquid crystals, we only need to consider the asymptotic interaction between two well separated single Skyrmions.

The configuration of a pair of Skyrmions, each described by the hedgehog ansatz, can be determined using a summation approximation. 
The asymptotics of such an approximation, following techniques from HEP \cite{Piette:1994ug} (details given in the Supplementary Information), shows that two Skyrmions with internal orientations $\gamma_A$ and $\gamma_B$ and separation $R$, have the effective interaction potential
\begin{align} \label{int_pot}
V_{\mbox{int}}\approx\frac{C^2\mu}{\pi J}\cos (\gamma_A-\gamma_B) \frac{e^{-\sqrt{\mu/J}R}}{\sqrt{R}},
\end{align}
where $\mu =|\bB|$ and $C$ is a fitting parameter. 
This interaction depends on the Skyrmions' relative internal orientation, $\gamma_A-\gamma_B$, suggesting in fact that Skyrmions attract when $|\gamma_A-\gamma_B| < \pi/2$. 
Statically, this is counteracted by the DMI which supports $\gamma_A=\gamma_B=-\frac{\pi}{2}$, hence Skyrmions usually repel. 
The potential also shows that Skyrmions interact less when $|\bB|=\mu$ is large; but this would give rise to sharply located Skyrmion which are potentially unstable.
This interaction can be used, in a pointwise manner, to model the interaction between an ensemble of Skyrmions as point-like particles.

\begin{figure}[!htb]
\includegraphics[width = 0.95\columnwidth]{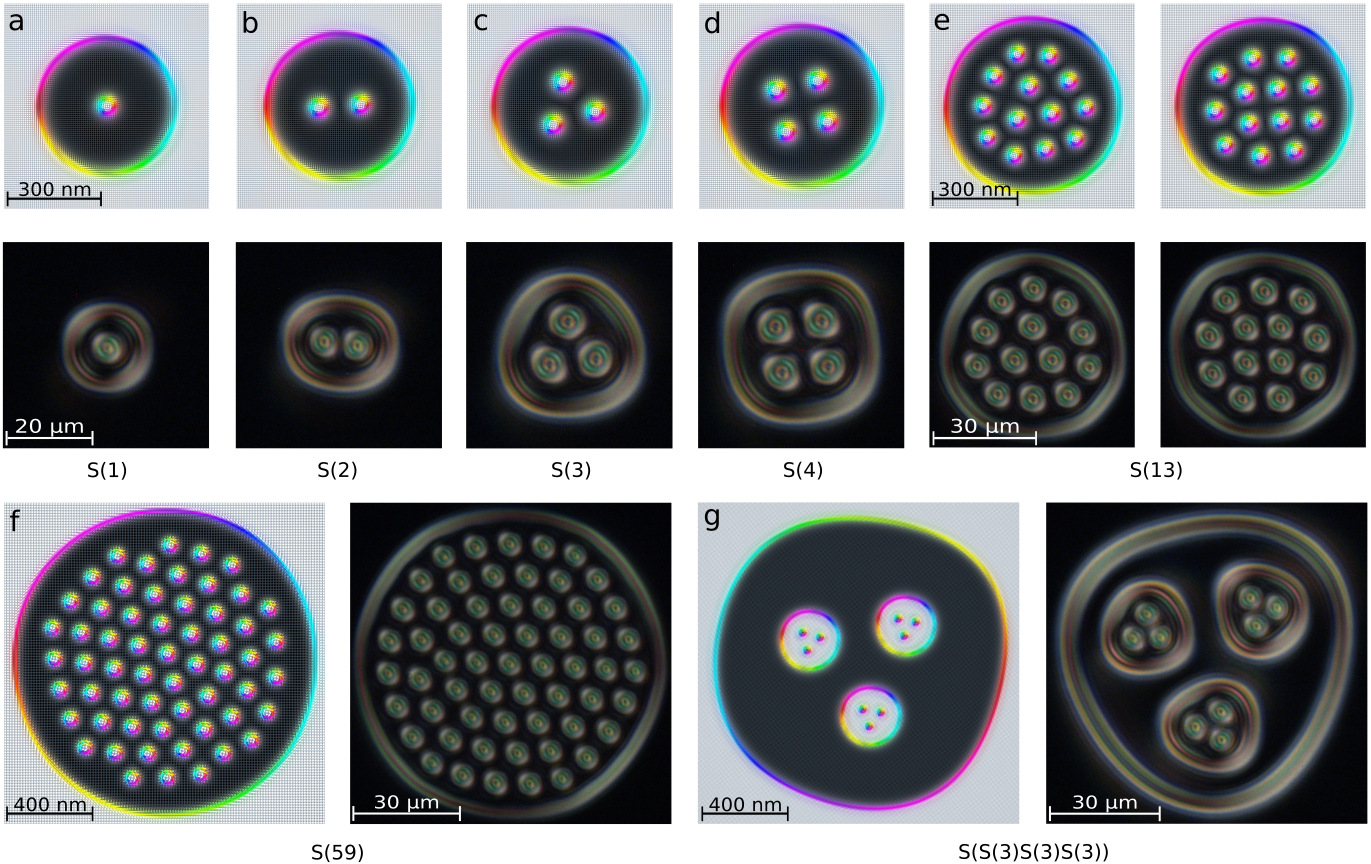}
\caption{\textbf{Skyrmion bag configurations.} 
   \textbf{(a-d)} Skyrmion bags $S(1)$ to $S(4)$, with magnetic MuMax3 simulations on the top row, and their experimental liquid crystal counterparts in the middle row. 
   The magnetic and liquid crystal bag configurations are clearly equivalent.
  \textbf{(e)} the two stable conformations of the $S(13)$ bag, which are analogous to the two best packings of 13 disks \cite{Fodor} (discussed in the Supplementary Information).
  Examples of more complex configurations are shown in \textbf{(f)} $S(59)$ bags, and \textbf{(g)} recursive $S(S(3)S(3) S(3))$ bags of topological degree 7.
}
\label{Sk_train}
\end{figure}

\begin{figure}[!htb]
  \includegraphics[width =0.95\textwidth]{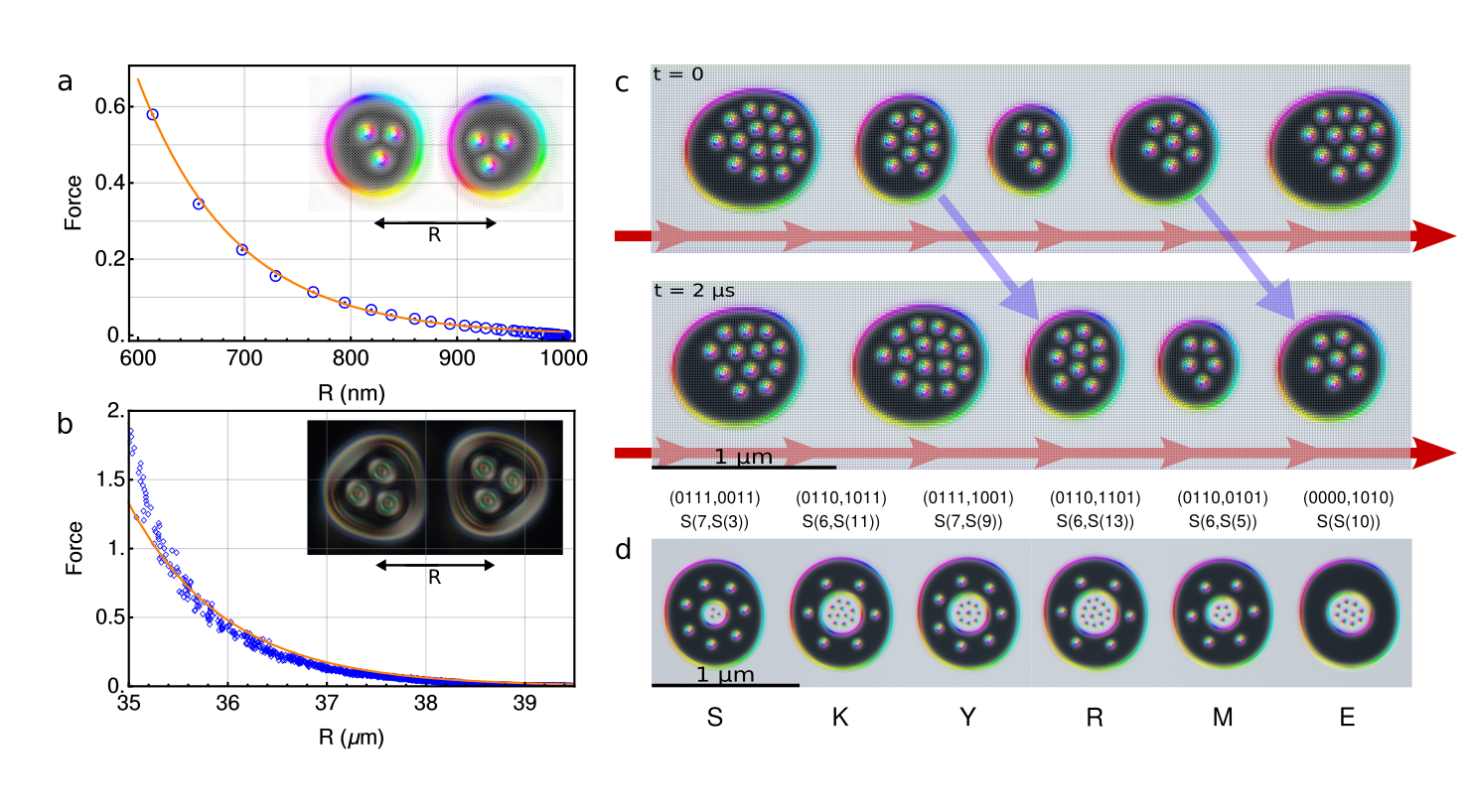}
   \caption{\textbf{Bag repulsion and Racetrack dynamics.}
   \textbf{(a)} Dynamical MuMax3 simulation of two $S(3)$ magnetic bags repelling (points), with the fitted dipole interaction potential \eqref{int_pot}. Typical $S(3)$ bag diameter in this simulation regime is $500$ nm.
   \textbf{(b)} Experimental measurement of two repelling $S(3)$ liquid crystal bags (points), with the fitted dipole interaction potential \eqref{int_pot} as in \textbf{(a)}.
   The interaction in these two cases has exactly the same form asymptotically as single Skyrmions. 
   \textbf{(c)} Two frames of a dynamical MuMax3 simulation of a collection of bags in a racetrack being pushed by a current.
   \textbf{(d)} A MuMax3 simulation of an encoding of the the word ``SKYRME" using a simple binary encoding of the alphabet where the outer bag holds the first four bits and the inner bag the next four bits and each full recursive bag represents a single letter and hence 8 bits of information.
 }
\label{msInt}
\end{figure}

We verified this interaction potential using micromagnetic simulations of two Skyrmions, using the MuMax3 finite-difference GPU accelerated code \cite{Mumax}. 
These simulations, incorporating the energy functional in equation (\ref{eng}) and nonlocal demagnetisation effects, resulted in the repulsive interaction force shown in figure \ref{msInt}\textbf{a}, which agrees with the force derived from the potential energy in equation \eqref{int_pot} (fitted for large separation). 
Experimental measurements of the repulsion between a pair of interacting Skyrmions in a chiral nematic liquid crystal is shown in figure \ref{Fig1}\textbf{j}.
Again, there is excellent agreement with a fit of the interaction $V_{\mbox{int}}$.

The form of our derived potential $V_{\mbox{int}}$ accounts for various previously observed phenomena.
Significantly, the orientation dependence explains the BiSkyrmion bound state \cite{BiSk} (each Skyrmion has opposite orientation), the orientations of magnetic bubbles \cite{Nag2}, and the relative orientation of Skyrmions in nano discs with respect to the boundary \cite{zhao2016direct}. 
Our analysis also shows that the force is weaker for large separation than previously known \cite{lin2013particle}.

The form of the potential in equation \eqref{int_pot} suggests that a concentric Skyrmion anti-Skyrmion ($m=-1$ in equation \eqref{anz}) configuration with a relative internal orientation of $\pi$, as in figure \ref{Sk_train}(a) might be stable. 
To create this configuration within MuMax3, we applied, in the region of the Skyrmion, a magnetic field opposite the external magnetic field, which caused it to grow. 
When the Skyrmion was sufficiently large, we turned off the magnetic field and then applied a second seeding magnetic field aligned with the external field, $\bB$. 
Although this nested Skyrmion configuration is not a global minimum energy configuration, it is a local minimum and is stable to large perturbations. 
We denote such `bag' configurations of $n$ anti-Skyrmions inside of a larger bag Skyrmion as $S(n)$, which has total topological degree $1-n$. 
Similarly, we created Skyrmion bags experimentally in chiral nematic liquid crystals using laser tweezers to stretch a Skyrmion to form the bag, and then used laser tweezers to place single Skyrmions inside of the bag. 
We repeated this process, for both magnetic simulations and experimental chiral nematic liquid crystals (up to $S(59)$) as shown in figure \ref{Sk_train}, we also produced nested bags \ref{Sk_train}(g).  Showing that liquid crystal and magnetic Skyrmion bags have similar configurations. 
In the chiral nematic liquid crystal experiments described here, $\bn$ is a director ($\bn=-\bn$), so there is no distinction between Skyrmions and anti-Skyrmions.

Skyrmion bags are the first example of stable composite Skyrmions in monochiral 2D materials.
Their range of possible configurations echoes the plethora of nuclei described by different topological degree configurations in the original Skyrme model. 
In particular the magnetic and liquid crystal configurations in figure \ref{Sk_train}(a), having total topological degree zero, resemble a sphaleron \cite{MantonSph} which plays a crucial role in nonperturbative calculations of electroweak theories.

Since isolated bags are circularly symmetric, we can apply the previous asymptotic interaction analysis (with different constants $C$). 
In figure \ref{msInt} the numerical and experimental results, for to interacting $S(3)$ bags, are compared with the previous interaction potential $V_{\mbox{int}}$, showing that they interact like large Skyrmions. 
Since static (anti-)Skyrmions repel, the packing of (anti-)Skyrmions inside a static bag can be mapped onto the mathematical circle packing problem \cite{Fodor} (discussed in the Supplementary Information).  
This also explains the packing of Skyrmions in nano discs \cite{zhao2016direct}.

In the proposed methods of racetrack memory, bits are encoded by Skyrmion separation.
The potential $V_{\mbox{int}}$ indicates that in fact Skyrmions in trains repel each other, eventually becoming uniformly distributed along the track, leading to data loss.
Devices utilising Skyrmion bags, on the other hand, will not have this problem, as $0,1$ and higher integers can be encoded in different integer $n$ in $S(n)$ bags, with no data  stored in their separation.
A train of Skyrmion bags, $S(n_1), S(n_2), \ldots$ for different integers $n_j$ storing bits, could be used as an alternative instead of a train of single Skyrmions, and the bags can be packed together and driven along by a current.
A full MuMax3 simulation of Skyrmion bags being accelerated along a race track, by a current, is shown in figure \ref{msInt}(c), where it can be seen that the inner anti-Skyrmions force the bags along.

There are a variety of ways a Skyrmion bag might be utilised to store data.
For instance, a Skyrmion bag could store 4 bits of data, by containing $1$ to $16$ Skyrmions, with about $220$nm of separation, requiring less than $1000$ nm (sizes of bags are considered further in the Supplementary information).
To encode 4 bits of data with single Skyrmions would need $1320$nm, hence bags could have a greater data density, as indicated in figure \ref{msInt}(c) to enocde the word `SKYRME'.

It is known that near a Skyrmion, the local magnetoresistance varies \cite{SkyDet}, and this phenomena can be read by tunnelling spin-mixing magnetoresistance (TXMR) using a non-magnetic scanning tunnelling microscope tip. 
This has the potential to `read' a Skyrmion and could be extended to read the contents of a Skyrmion bag. 
A device could be made from an ultrathin metallic ferromagnetic material \cite{Woo2016}, where Skyrmions have been previously formed and experimentally driven. 
Skyrmions could be formed using an electric potential from a tip (previously realised experimentally \cite{hsu}), and the Skyrmions could then be forced to grow using a fine magnetic field. 
The inner bag antiSkyrmions could then be formed, within the larger Skyrmion, by an electric field. 
A train of bags could then be pushed with a current (as in our MuMax3 simulation in figure \ref{msInt}) and read with a TXMR head \cite{SkyDet}.
But here, due to the differing bag contents, each site could encode more than a zero or a one, further increasing data density. 
Such a Skyrmion bag racetrack device would further mitigate drawbacks of conventionally proposed racetrack memory, since repulsion or sticking on defects would not lead to data loss.

In summary, our theory of interactions demonstrates an analogues interaction between Skyrmions in HEP, magnetic systems and liquid crystal films. 
Not only does the interaction's orientation-dependence describe previous observations \cite{BiSk, yu2010real,Nag2,zhao2016direct,lin2013particle}, it has also led to the formulation of Skyrmion bags which opens up new possibilities for composite Skyrmion excitations in materials.  
Magnetic Skyrmion bags have the potential to extend Skyrmionics beyond binary bits, suggesting Skyrmion racetrack memory devices with a significantly higher possible storage density. 
More generally our overall approach hints at new physics in magnetic and liquid crystal motivated by theoretical approached in HEP.

\section{Methods}
\subsection{Micromagnetic Simulations}
Throughout we performed magnetic simulations using the GPU-accelerated micromagnetic simulation program MuMax3 \cite{Mumax} with Landau-Lifshitz dynamics in the form,
\begin{align}
\frac{\partial \mathbf{n}}{\partial x}=\hat{\gamma}\frac{1}{1+\alpha^2}\left(\mathbf{n} \times \mathbf{B}_{\mathrm{eff}}+\alpha\mathbf{n}
\times\left( \mathbf{n} \times \mathbf{B}_{\mathrm{eff}} \right) \right),
\end{align}
where $\hat{\gamma}$ is the gyromagnetic ratio, $\alpha$ the dimensionless damping parameter, $\mathbf{B}_{\mathrm{eff}}$ the efffective field and $\mathbf{n}$ the magnetisation unit vector. The simulations where performed with free boundary conditions.  The effective field energy includes contributions from exchange, anisotropy, DMI and applied field.

The simulation geometry is typically a $1024\times1024$ nm$^2$ square of $1$ nm thickness with a cell size of $2 \times 2\times$ 1 nm$^3$ although for finer detail cell sizes of $1$ nm$^3$ where used. Material parameters are uniformly throughout all simulations: saturation magnetisation $M_{sat} = 900$ kAm$^{-1}$, exchange $A = 10$ pJm$^{-1}$, damping $\alpha = 0.5$, DMI $D = 0.6$mJm$^{-2}$ and uniaxial anisotropy $K_{u} = 0.56$MJm$^{-3}$. The anisotropy is along the $+z$ direction. The material parameters where taken from Woo et al.  \cite{Woo2017} and tuned to maximise the stability of the Skyrmion bags.

\subsection{Liquid crystal experiment}
\textbf{Materials and sample preparation.}\\
To assure accessibility and a broad impact of our work, pentylcyanobiphenyl (5CB, from EM chemicals), a commonly used and commercially available nematic liquid-crystal (LC) material, was doped with small amounts of chiral additive, cholesterol pelargonate  (Sigma-Aldrich), resulting in a chiral nematic LC. The helicoidal pitch $p$ of the LC mixture is determined by $p=(\xi\cdot c)^{-1}$  where $c$ is the weight fraction of the additive and $\xi=6.25\mu$m$^{-1}$, the helical twisting power of the additive in 5CB. Confining glass substrates were treated with polyimide SE1211 (obtained from Nissan Chemicals) to ensure vertical alignment of liquid crystal orientation at the LC-glass interface. Polyimide was applied to substrates by spin-coating at 2700 rpm for 30 s and then baked for 5 min at $90^o$C then 1 h at $180^o$C. LC cells with gap thickness of $d=10-20 \mu$m were produced by sandwiching glass fiber segments in UV-curable glue. In cells where $d/p\approx1$, spontaneous and controllably generated structures corresponding to minima of free energy were generated and manipulated using laser tweezers, as detailed below.

\textbf{Optical generation of Skyrmions and Skyrmion bags.}\\
Holographic laser tweezers capable of producing arbitrary patterns of laser light intensity within the LC sample based on an Ytterbium-doped fiber laser (YLR-10-1064, IPG Photonics, operating at 1064 nm) and a phase-only spatial light modulator (P512-1064, Boulder Nonlinear Systems) integrated on an inverted microscope (IX81, Olympus) enabled generation of Skyrmions and Skyrmion bags and subsequent manipulations. Polarizing optical microscopy and video microscopy observations of LC Skyrmions was achieved with a charge coupled device camera (Grasshopper, PointGrey Research) \cite{Fuk}.

\subsection{Data availability}
The datasets generated during and/or analysed during the current study are available from the corresponding author on reasonable request.


\textbf{Acknowledgements} \\
D.F. and M.R.D. acknowledge the fund by the Leverhulme Trust Research Programme Grant RP2013-K-009, SPOCK: Scientific Properties Of Complex Knots. DF also thanks P.M. Sutcliffe, M. Gradhand, A. Bogdanov and W. Zakrzewski for comments. Experimental research at CU-Boulder (P.J.A., J-S. T. and I.I.S.) was supported by the U.S. Department of Energy, Office of Basic Energy Sciences, Division of Materials Sciences and Engineering, under Award ER46921, contract DE-SC0010305.

\textbf{Author contributions}
C.K. performed the numerical analysis and D.F. performed the theoretical analysis, with input from C.K. and M.R.D. C.K. and D.F. proposed the investigation.
P.J.A. and J-S.T. performed the experiment with suggestions from I.I.S. and D.F. I.I.S. provided experimental techniques and materials. P.J.A., J-S.T. and I.I.S analysed experimental data. D.F., C.K. and M.R.D. prepared the manuscript with input from all authors.

\newpage
\section{Supplementary Information}
\beginsupplement
\setcounter{equation}{0}
\renewcommand{\theequation}{S\arabic{equation}}
\textbf{Magnetic and chiral nematic liquid crystals:}
The link between magnetic and chiral nematic liquid crystal Skyrmions can be seen from the Frank free energy,
\begin{align}
E=\int \left(\frac{K_{11}}{2}(\nabla\cdot\bn)^2+ \frac{K_{22}}{2}(\bn\cdot(\nabla \times \bn))^2 +
 \frac{K_{33}}{2}(\bn\times(\nabla\times\bn)) +q_0 K_{22}\bn\cdot(\nabla\times\bn) \right)d^2 \bx,\nonumber
\end{align}
where on setting the bend, twist and splay energies to be equal, $J=K_{11}=K_{22}=K_{33}$, $D=K_{22}q_0$, and the pitch axis to $q_0=\frac{2\pi}{p}$, and including an external magnetic field $\bB$, we derive the energy functional \eqref{eng}, of the particularly successful magnetic Skyrmion model \cite{nagaosa2013topological} for the spin field $\bn(\bx)$.

%

\textbf{Minimum energy:} The minimum energy Skyrmion configuration has been well studied \cite{melcher2014chiral}, and it is a solution of the energy functional's corresponding Euler-Lagrange equations,
\begin{align}\label{EQM}
J\pa_{ii}\bn-2D\nabla \times \bn +\bB =0.
\end{align}
The dot product of \eqref{EQM} with $\bn$ can be used to express the energy functional as the interaction of $\bn$ with the external magnetic field,
\begin{align}
E=-\frac{1}{2}\int \bB\cdot \bn ~ d^2x. 
\end{align}
It must be stressed that this expression can only be used to calculate the energy of a configuration, $\bn$, which is a solution of \eqref{EQM}.

\textbf{Minimiser over all homotopy classes:}
On substitution the hedgehog ansatz separates angular and radial components in the energy functional \eqref{eng}. 
The angular component can then be integrated out, giving the radially dependent energy functional,
\begin{align}\label{engrA}
E=&2\pi \int \left( \frac{1}{2}f'^2(r)+\frac{m^2}{2r^2}\sin^2 f(r) \right.\nonumber \\ 
&+  D\frac{\sin(m\pi)\sin(\gamma+m\pi)}{m-1}\left(f'+\frac{m}{2r}\sin \left(2 f(r)\right)\right) \nonumber\\
&-\left.\mu \cos f(r)\right)r dr.
\end{align}
Using a scaling argument \cite{derrick1964comments}, $r\to\lambda r$, it can be seen that there are no static solutions unless $\frac{\sin(m\pi)\sin(\gamma+m\pi)}{m-1}$ is non-zero. 
Hence, there are only stable solutions for $m=1$ Skyrmions, a concise and detailed discussion of magnetic Skyrmion stability can be found in \cite{melcher2014chiral}.

\textbf{Finding the minimum energy axial solution:}  In order to derive the minimum energy solution of \eqref{engrA}, we take its functional derivative,
\begin{align}\label{eqmr}
&Jrf''(r)+Jf'(r)+D\sin(\gamma)\left(1-\cos(2f(r)\right)  \nonumber\\
&-\frac{J}{2r}\sin\left(2f(r)\right)-\mu r \sin\left(f(r)\right)=0,
\end{align}
and then perform a numerical gradient flow on a lattice of $1000$ points, with the boundary conditions $f(0)=\pi$, $f(\infty)=0$, shown in figure \ref{f1-subplot}. This is the techniques used to find the one dimensional solutions through the paper.

\textbf{Approximating Skyrmions as dipoles.}  Importantly, asymptotically to first order, the Dzyaloshinskii\hyp{}Moriya interaction becomes negligible for the circularly symmetric Skyrmion. Hence far from the core of the Skyrmion, $f(r)$ is a solution of the differential equation,
\begin{align}\label{fasym}
f''(r)+\frac{1}{r}f'(r)-(\frac{1}{r^2}+\mu_J)f(r)=0,
\end{align}
where $\mu_J=\mu/J$ which is defined for convenience.

This differential equation is solved by a modified Bessel function of first order, $K_1$, which can readily be approximated asymptotically as an exponential. Therefore, as $r\to\infty$ we find,
\begin{align}\label{Eq.:Assympt}
f &\sim \frac{C \sqrt{\mu_J} }{2\pi}K_1(\sqrt{\mu_J} r)\sim C \frac{e^{\sqrt{\mu_J}r}}{\sqrt{r}}\ \text{,}
\end{align}
where $C$ is a constant which depends on both $J$ and $\mu$, and is found by fitting to the exponential tail of $f(r)$, as in figure \ref{f1-subplot}. 
We further simplified the asymptotic form in Eq.~(\ref{Eq.:Assympt}) exploiting the asymptotic behaviour of $K_1(\sqrt{\mu_J}r)$, and such an asymptotic tail is also shown in figure \ref{f1-subplot}.

From this we can define the asymptotic form of $\bn(\bx) \approx \be + \bpsi(\bx)$ as (where $\be=(0,0,1)$),
\begin{align}
\bpsi(\bx)=\frac{C\sqrt{\mu_J}}{2\pi}K_1(\sqrt{\mu_J}r) (\cos(\theta+\gamma),\sin(\theta+\gamma) ,0). \nonumber
 \end{align}
If we introduce the orthogonal dipoles,
\begin{align}
\bp_1=C\left(\cos\gamma,-\sin\gamma\right), ~~ \bp_2=C\left(\sin\gamma,\cos\gamma\right),  \nonumber
\end{align}
we can rewrite the asymptotic form, $\bpsi(\bx)=(\psi_1(\bx),\psi_2(\bx),0)$, as
\begin{align}\label{dipole}
\psi_a(\bx)=\frac{\sqrt{\mu_J}}{2\pi}\bp_a\cdot \hat{\bx}K_1(\sqrt{\mu_J}r)=-\frac{1}{2\pi}\bp_a\cdot \nabla K_0(\sqrt{\mu_J}r),
\end{align}
where $a=1,2$.\\
Hence, far from a single Skyrmions the fields behave as a pair of orthogonal dipoles.

\textbf{Interaction potential of two Skyrmions:}
In order to understand how Skyrmions interact we analytically construct a two Skyrmion configuration.
In Ref.~\cite{Piette:1994ug} it was shown that a configuration of two well separated Skyrmions can be approximated by,
\begin{align}
W_{\rm{config}}=W_{1}(\bx)+W_{2}(\bx).  \nonumber
\end{align}
where $W_{1}$ and $W_{2}$ are the Skyrmion configurations $\bn$ and $\bmm$ represented in stereographic coordinates,
\begin{align}
W_1(\bx)=\frac{n_1(\bx)+in_2(\bx)}{1+n_3(\bx)},  ~W_2(\bx)=\frac{m_1(\bx)+im_2(\bx)}{1+m_3(\bx)}.\nonumber
\end{align}
We can use this to located one Skyrmion at the origin and a second at $\bR$.

\begin{figure}[!htb]
\includegraphics[width = 0.9\columnwidth]{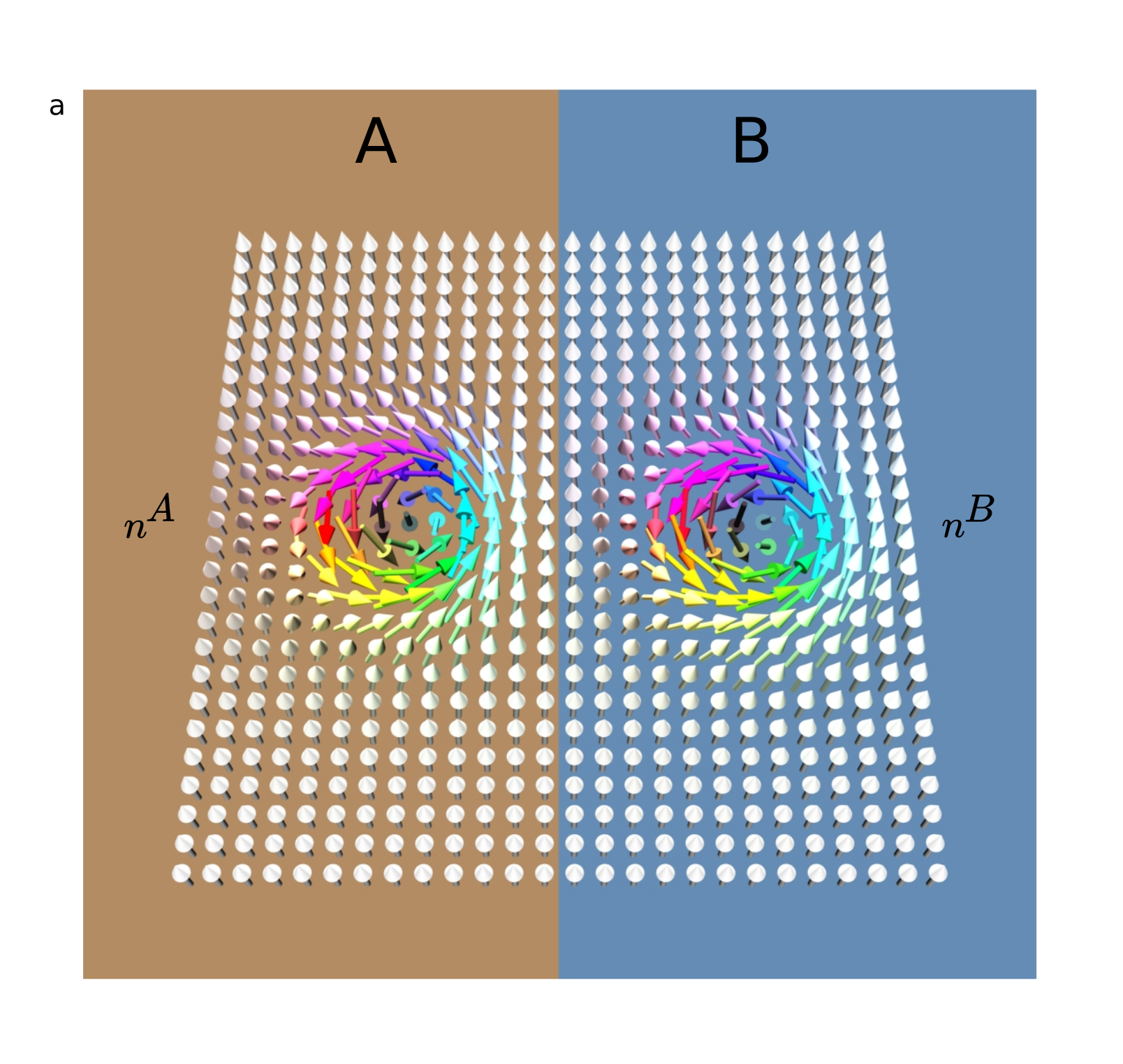}
\caption
{\textbf{A two Skyrmion configuration.}
\textbf{a,} A two Skyrmion configuration split into regions. Region $A$ contains $\bn^A$ and region $B$ contains $\bn^B$.
}
\label{2_Sk_reions}
\end{figure}

\begin{figure}[!htb]
\includegraphics[width = 0.8\columnwidth]{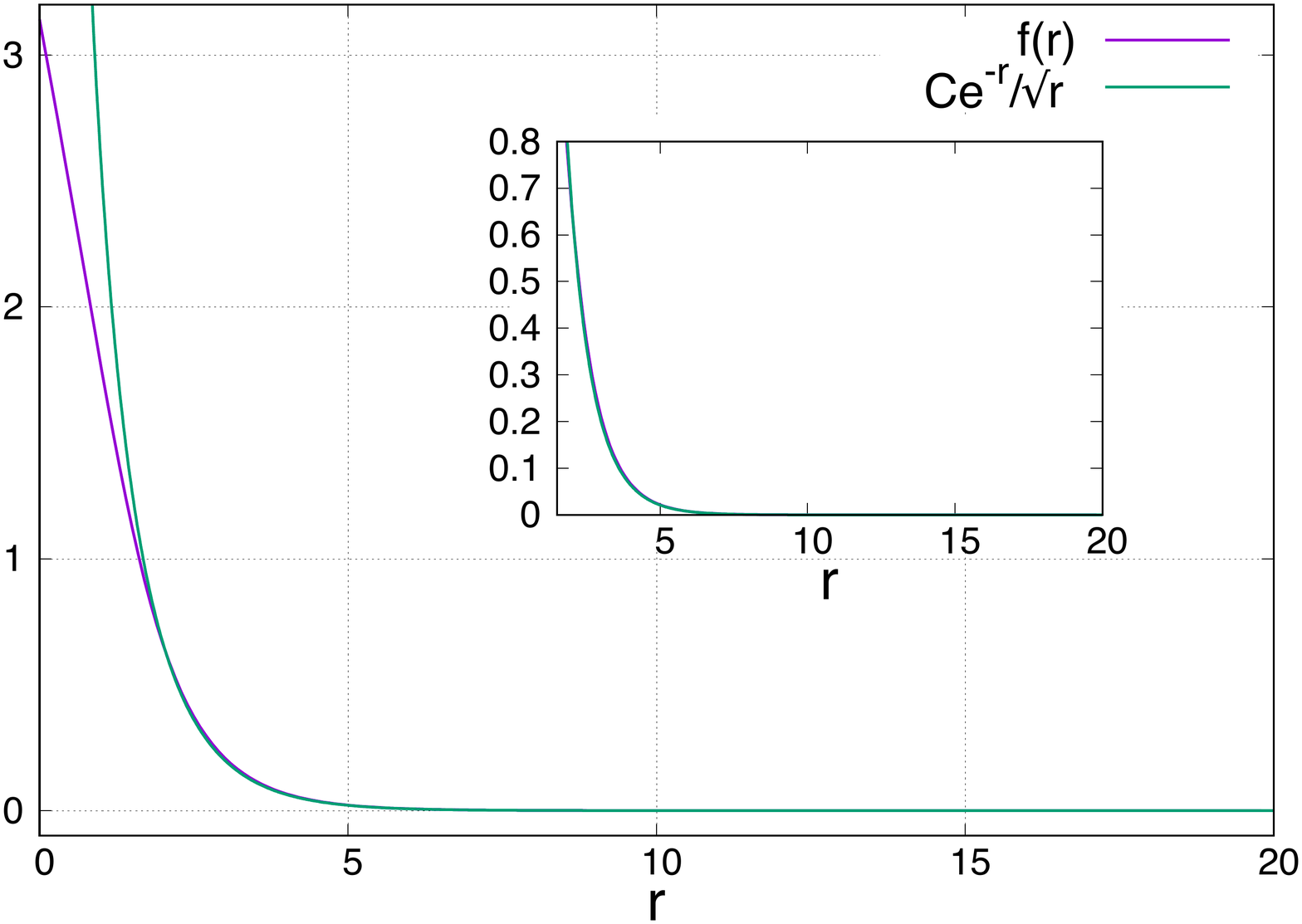}
\caption{\textbf{Minimum energy profile.} The minimum energy Skyrmion profile function, $f(r)$, for $D=J=\mu=1$ (chosen for aesthetics) and $\gamma=-\pi/2$ of the circularly symmetric hedgehog ansatz \eqref{anz}. Also plotted is the exponential tail approximation \eqref{Eq.:Assympt}, where $C\approx6.95$ is found to be optimal. The tail approximation is a good approximation for $r>2$ and is used in the asymptotic interaction potential.}
\label{f1-subplot}
\end{figure}

The vector valued configuration, $\bn(\bx)$, can then be recovered by inverse stereographic projection and produces the configuration shown in figure \ref{2_Sk_reions}. This summation ansatz can be extended to produce a configuration of any number of Skyrmions by simply adding more configurations together.

We will consider a configuration of two well separated hedgehog Skyrmions, as in figure \ref{2_Sk_reions}, where one Skyrmion $\bn^A$ is in region $A$ and the second Skyrmion $\bn^B$ is in region $B$. In region $A$ the Skyrmion $\bn^B$ is close to the boundary value $\be$ and in region $B$ the Skyrmion $\bn^A$ is close to the boundary value $\be$. This means we can approximate the configuration $\bn^A$ in region $B$ and configuration $\bn^B$ in region $A$ as,
\begin{align}
\bn^A =\sqrt{1-\bpsi^A\cdot \bpsi^A}\be \approx \be + \bpsi^A ~\mbox{and} ~\bn^B =\sqrt{1-\bpsi^B\cdot \bpsi^B}\be \approx \be + \bpsi^B,
\end{align}
where $\bpsi^A\cdot \be =\bpsi^B\cdot \be=0$. It can be shown, \cite{Piette:1994ug}, using the summation ansatz, that a two Skyrmion configuration, $\bn^{AB}$, with a large separation, $R=|\bR|$, in region $A$ can be approximated as
\begin{align}
\bn^{AB} \approx \bn^A +\bvarepsilon^B\times\bn^A+\frac{1}{2}\bvarepsilon^B\times(\bvarepsilon^B\times\bn^A),
\end{align}
and in region $B$ as
\begin{align}
\bn^{AB} \approx \bn^B +\bvarepsilon^A\times\bn^B+\frac{1}{2}\bvarepsilon^A\times(\bvarepsilon^A\times\bn^B),
\end{align}
where,
\begin{align}
\bvarepsilon^B = \frac{1}{2}\bn^A\times\left((1+\bn^A\cdot \be)\bpsi^B-(\bn^A\cdot\bpsi^B)\be\right), \\
\bvarepsilon^A = \frac{1}{2}\bn^B\times\left((1+\bn^B\cdot \be)\bpsi^A-(\bn^B\cdot\bpsi^A)\be\right).
\end{align}
We can exploit this to approximate the energy of a two Skyrmion configuration, $\bn^{AB}$, as the sum of the contributions over the the two regions $A$ and $B$ as in figure \ref{2_Sk_reions}. This gives the energy functional,

\begin{align}\label{Esum}
E[\bn^{AB}]\approx& 2 E_{\mbox{Skyrmion}} \\ 
+&\int_B d^2x\left(\frac{J}{2}\pa_i \bpsi^A\cdot\pa_i \bpsi^A +\frac{1}{4}D\bpsi^A\cdot \nabla \times \bn^A +\frac{1}{2} \mu \bpsi^A\cdot \bpsi^A\right) \nonumber \\ 
+&\int_A d^2x\left(\frac{J}{2}\pa_i \bpsi^B\cdot\pa_i \bpsi^B +\frac{1}{4}D\bpsi^B\cdot \nabla \times \bn^B +\frac{1}{2}\mu \bpsi^B\cdot \bpsi^B\right) \nonumber \\
+&\int_A \bvarepsilon^B \cdot \left(J\bn^A\times\pa_i^2\bn^A+2D\bn^A\times\nabla\times\bn^A+\bB\times\bn^A\right) \nonumber \\
+&\int_B \bvarepsilon^A \cdot \left(J\bn^B\times\pa_i^2\bn^B+2D\bn^B\times\nabla\times\bn^B+\bB\times\bn^B\right).\nonumber
\end{align}
The first two integrals are the energy of the Skyrmion tails in the complementary empty region, and the second two integrals measure the overlap of one Skyrmion with the other.
We can evaluate this integral, for two Skyrmions with a large separation, by linearising the above integral over the area $A$ and keeping only first order terms in $\bpsi^A$ and $\bpsi^B$. This gives,
\begin{align}\label{lim_eq}
\int_A(\bpsi^B\cdot(J\nabla^2-\mu)\bpsi^A +D\bpsi^B\cdot\nabla\times\bpsi^A)d^3x.
\end{align}
The fact that $\bpsi^A$ and $\bpsi^B$ are normal to $\be$ and that $\nabla \times \bpsi^A$ is parallel with $\be$ implies that $\bpsi^B\cdot\nabla\times\bpsi^A\approx0$.
From the previous asymptotic analysis, we can see that the asymptotic form of $(\nabla^2-\mu_J)\bpsi^A=0$. We can then use the Green function, $K_0(\sqrt{\mu_J}r)$, of the 2D Klein-Gordon equation,
\begin{align}
(\nabla^2-\mu_J)K_0(\sqrt{\mu_J}r)=-2\pi\delta^{(2)}(\bx),
\end{align}
to re-express the asymptotic field of $\bn^a$ in the integral \eqref{lim_eq} as a product of orthogonal dipoles,
\begin{align}
\left( \nabla^2-\mu_J\right)\bpsi^A_a(\bx)=\bp_a\cdot\nabla\delta^{(2)}(\bx), ~~a=1,2.
\end{align}
From equation \eqref{dipole} we know the asymptotic form of the second Skyrmion, $\bn^B$, at $R$ is,
\begin{align}
\bpsi^B_a=\frac{1}{2\pi}\bp_a(\gamma^B)\cdot\nabla K_0 (\sqrt{\mu_J}|\bx-R|), ~~a=1,2.
\end{align}
Combining these results we can evaluate the interaction potential defined as $V_{\mbox{int}}=E[\bn^{AB}]-E[\bn^A]-E[\bn^B]$,
\begin{align}
V_{\mbox{int}}=&J\int_A d^2x\bpsi^B\cdot(\nabla^2-\mu_J)\bpsi^A \nonumber \\
=&\frac{C^2\mu}{\pi}\cos \gamma K_0(\sqrt{\mu_J}R),
\end{align}
where $\gamma=\gamma^A-\gamma^B$.
We find that two well separated Skyrmions interact as a pair of orthogonal dipoles. Using the asymptotic form of $K_0$ the interaction takes the form 
\begin{align}
V_{\mbox{int}}\approx\frac{C^2\mu_J}{\pi}\cos \gamma \frac{e^{-\sqrt{\mu_J}R}}{\sqrt{R}},
\end{align}
where $\mu =|\bB|$, $R$ is the inter-Skyrmion separation and $C$ is a constant found by fitting the asymptotic tail of $f(r)$. \\

\textbf{Orientation of Skyrmion inside of the bag.}
In terms of the hedgehog ansatz, the usual boundary conditions are $f(0)=\pi, f(\infty)=0$. For the single Skyrmion bag $f(0)=0, f(\infty)=0$. As we can see from the images, we can divide the space into two regions, the first region $r<R_0$ contains the anti-Skyrmion and the second contains the Skyrmion where $f(R_0)=\pi$. Considering the central region for large $R_0$ then we can transform the profile function as $f \to -f+\pi$, this transformation causes the DMI contribution to pickup a minus sign and hence $\gamma=\frac{\pi}{2}$ is the stable configuration.\\

\textbf{Circle packing} The packing of Skyrmions inside of a bag can be mapped onto the mathematical circle packing problem \cite{Fodor}, and is some times referred to as the inimical dictators problem. Circle packing is concerned with the configurations of packing the largest equal size circles into the area of a disk, qualitatively the configurations match the inner Skyrmions of the bags in both magnetic and chiral nematic liquid crystal cases. This is because, as the interaction potential shows, they repel each other exponentially with distance and hence each Skyrmion occupies a given circular area, similar to the circle packing problem. Importantly, it has been shown that packing thirteen circles into a disk has two optimal configurations, \cite{Fodor}. This corresponds to the $S(13)$ configuration, where we also find two stable configurations, for both magnetic and chiral nematic liquid crystals, which correspond to the two circle packing configurations and they are shown in figure \ref{Sk_train}.

This shows that if a magnetic $S(13)$ bag configuration was experimentally produced, data could be stored by the two configurations, where the configuration could be switched from one confirmation to the other. This would alleviate the problem of Skyrmion repulsion affecting Racetrack memory and the possibility of a single Skyrmion `sticking' on a defect and effecting the encoded data.

\textbf{Size of Skyrmion bags.}
Each additional Skyrmion inside the Skyrmion bag increases the radius of the bag due to the repulsion between internal Skyrmions and between those Skyrmions and the bag boundary. We find this relationship to be very close to linear up to the $S(15)$ bag.
\begin{figure}[!htb]
\includegraphics[width = 0.8\columnwidth]{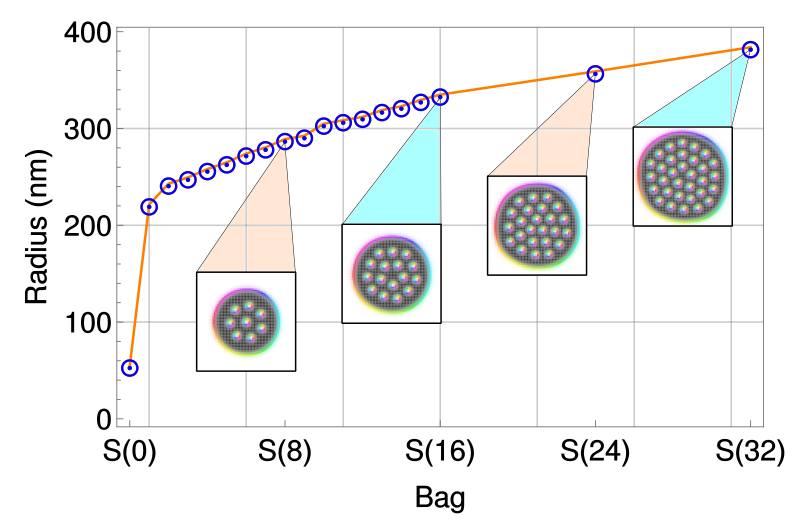}
\caption{\textbf{Magnetic Skyrmion Bag radius.} Mumax3 Skyrmion bag simulations run on a fixed $1024$ nm square domain configured as outlined in the Methods section. The $S(n)$ bags are relaxed and then measured from the widest outer domain wall boundary to show how their radii increase with increasing $n$. $S(0)$, in this terminology is a standard, charge one, Skyrmion.}
\label{MagBagRad}
\end{figure}

\begin{figure}[!htb]
\includegraphics[width = 0.8\columnwidth]{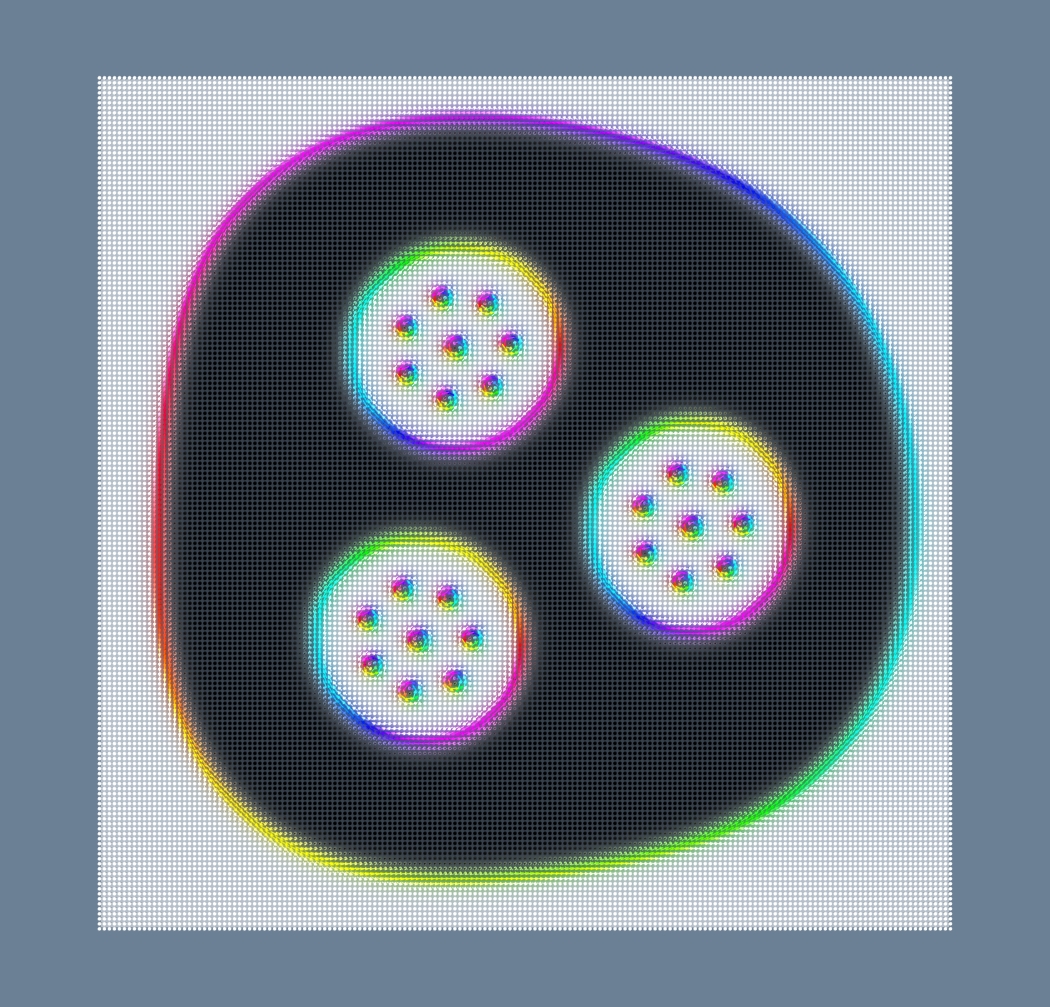}
\caption{\textbf{A magnetic $S(S(8)S(8)S(8))$ bag simulation, of topological degree $22$.}
A stable, recursive Skyrmion bag, Mumax3 simulation run on a fixed $2048$ nm square domain configured as outlined in the Methods section.
}
\label{MagBag7}
\end{figure}

\end{document}